\documentclass[%
reprint,
showpacs,preprintnumbers,
amsmath,amssymb,
longbibliography
]{revtex4-2}
\usepackage{graphicx}
\usepackage{dcolumn}
\usepackage{bm}
\usepackage{colortbl}
\usepackage[unicode=true,colorlinks=true,citecolor=blue,urlcolor=blue]{hyperref}
\usepackage{braket}
\usepackage[normalem]{ulem}

\let\ifr\i
\renewcommand{\i}{{\rm i}}
\renewcommand{\d}{\mathrm d}
\renewcommand{\emph}{\textit}
\renewcommand{\braket}[1]{\left\langle #1 \right\rangle}
\newcommand{\eps}{\varepsilon}

\begin{document}

	\title{Modification of Hanle and polarization recovery curves \\under interplay of hopping and quantum measurement back action}

	\author{A.~L.~Zibinskiy}
	\email[Electronic address: ]{zibinskiy.coherent@mail.ioffe.ru}
	\affiliation{Ioffe Institute, 194021 St. Petersburg, Russia}
	\author{D.~S.~Smirnov}
	\affiliation{Ioffe Institute, 194021 St. Petersburg, Russia}
	
	\begin{abstract}
		The measurements of Hanle and polarization recovery effects for localized charge carriers are the basic tools for determining parameters of the spin dynamics, such as strength of the hyperfine interaction, for example, in quantum dots. We describe the dependence of the spin polarization of localized electrons on transverse and longitudinal magnetic fields taking into account the interplay between electron hopping and measurement back action. We show that these two have a qualitatively similar effect in the Faraday geometry, but compete in the Voigt geometry. This allows one to describe a broad range of the experimental results and study the fundamental effects of quantum measurements.
	\end{abstract}
	
	\maketitle{}
	
	\section{Introduction}
	\label{sec:Intro}
	
	The discovery of optical orientation in semiconductors made by Lampel in 1968~\cite{lampelNuclearDynamicPolarization1968} initiated the intensive study of spin physics in the solid state. Since then, a tremendous progress has been made jointly in theory, experiment and technology, and thus the research area of semiconductor spintronics was born~\cite{dyakonov_book}. Nowadays it is impossible to overestimate the fundamental importance of the obtained results and their role for the prospects of quantum information processing~\cite{burkardSemiconductorSpinQubits2023}.
	
	In particular, the spin physics of the localized charge carriers attracts a lot of attention. Partially, this is because the localized spins represent a natural two-level system for the role of a stationary qubit~\cite{LossPRA98}. Additionally, they have very long spin relaxation times due to the suppression of the spin relaxation mechanisms related to the orbital electron motion~\cite{Elliott54,yafet63,DP,dyakonov72,BAP}. Instead the dynamics and relaxation of localized electron spins is driven by the hyperfine interaction with the host lattice nuclei~\cite{GlazovBook2018}.
	
	The basic methods of optical initialization~\cite{gerardot08}, manipulation~\cite{press08} and readout~\cite{berezovsky2006nondestructive} of electron spins are already well established. A very simple and useful method for experimental determination of the spin relaxation time is the measurement of the Hanle effect, which that is the suppression of the spin polarization in transverse magnetic field~\cite{hanleUeberMagnetischeBeeinflussung1924}. For localized electrons, there is also a complementary phenomenon of polarization recovery in a longitudinal magnetic field, which provides additional information about spin relaxation and hyperfine interaction~\cite{PhysRevLett.88.256801,smirnovSpinPolarizationRecovery2020}.
	
	Hanle and polarization recovery curves were measured experimentally for a great deal of different systems with localized electrons:
	Donor-bound electrons in bulk semiconductors~\cite{PhysRevLett.88.256801,heisterkampLongitudinalTransverseSpin2015,bussElectronSpinDynamics2016,sokolovNuclearSpinCooling2017,kirsteinExtendedSpinCoherence2021,kudlacikOpticalSpinOrientation2024},
	quantum wells and other heterostructures~\cite{koptevaSpinDephasingElectrons2019, kenEffectElectricCurrent2020},
	and self-assembled quantum dots~\cite{dzhioevStabilizationElectronNuclearSpin2007,auerMeasurementKnightField2009,mikhailovElectronHoleSpin2018, rautertOpticalOrientationAlignment2019,nekrasovEffectNuclearQuadrupole2019}
	are just a few examples.
	Starting from the pioneering work of Merkulov, Efros and Rosen~\cite{Merkulov2002}, a number of theoretical models of the Hanle and polarization recovery effects in quantum dots have been developed. It was found that the anisotropy of the hyperfine interaction~\cite{MX2_Avdeev,PhysRevB.101.075412} and repetition period of the pump and probe pulses~\cite{efros03,yugova12,PhysRevB.98.125306,PhysRevResearch.1.033189} may play a role for the shape of the spin polarization dependence on transverse and longitudinal magnetic field. Also a particularly important effect is often the finite nuclear spin correlation time~\cite{schulten,Glazov_hopping,PhysRevLett.134.016201} which breaks the assumption of the static nuclear spin fluctuations~\cite{Merkulov2002}. Still, many experiments cannot be fully explained by any existing model.
	
	We suggest, that this may be partially related to the effect of the quantum measurement back action, which unavoidably takes place in any experiment. In the case of localized electrons, it leads to the suppression of the transverse spin components under the action of a probe (or pump) beam~\cite{PhysRevB.103.045413,Leppennen2022,Zeno_exp}. According to the general principles, this may lead to the quantum Zeno~\cite{khalfin1958contribution,doi:10.1063/1.523304} and anti-Zeno~\cite{LANE1983359,Facchi_2008} effects, which modify the spin dynamics, including parameters of Hanle and polarization recovery curves.
	
	
	The paper is organized as follows. In Sec.~\ref{sec:Model} we present a theoretical model for the spin dynamics and solve it for the steady state. In Sec.~\ref{sec:Results} we describe the parameters of polarization recovery and Hanle curves as functions of the hopping rate and the measurement strength and present simple analytical expressions for a few limiting cases. Sec.~\ref{sec:Discussion} contains discussion of our results in relation to the experiments, including violation of some of the theoretical assumptions. A brief summary of the results is given in Sec.~\ref{sec:Conclusion}.

	\section{Model}
	\label{sec:Model}
	
\begin{figure}[htpb]
		\begin{center}
			\includegraphics[width=1.0\linewidth]{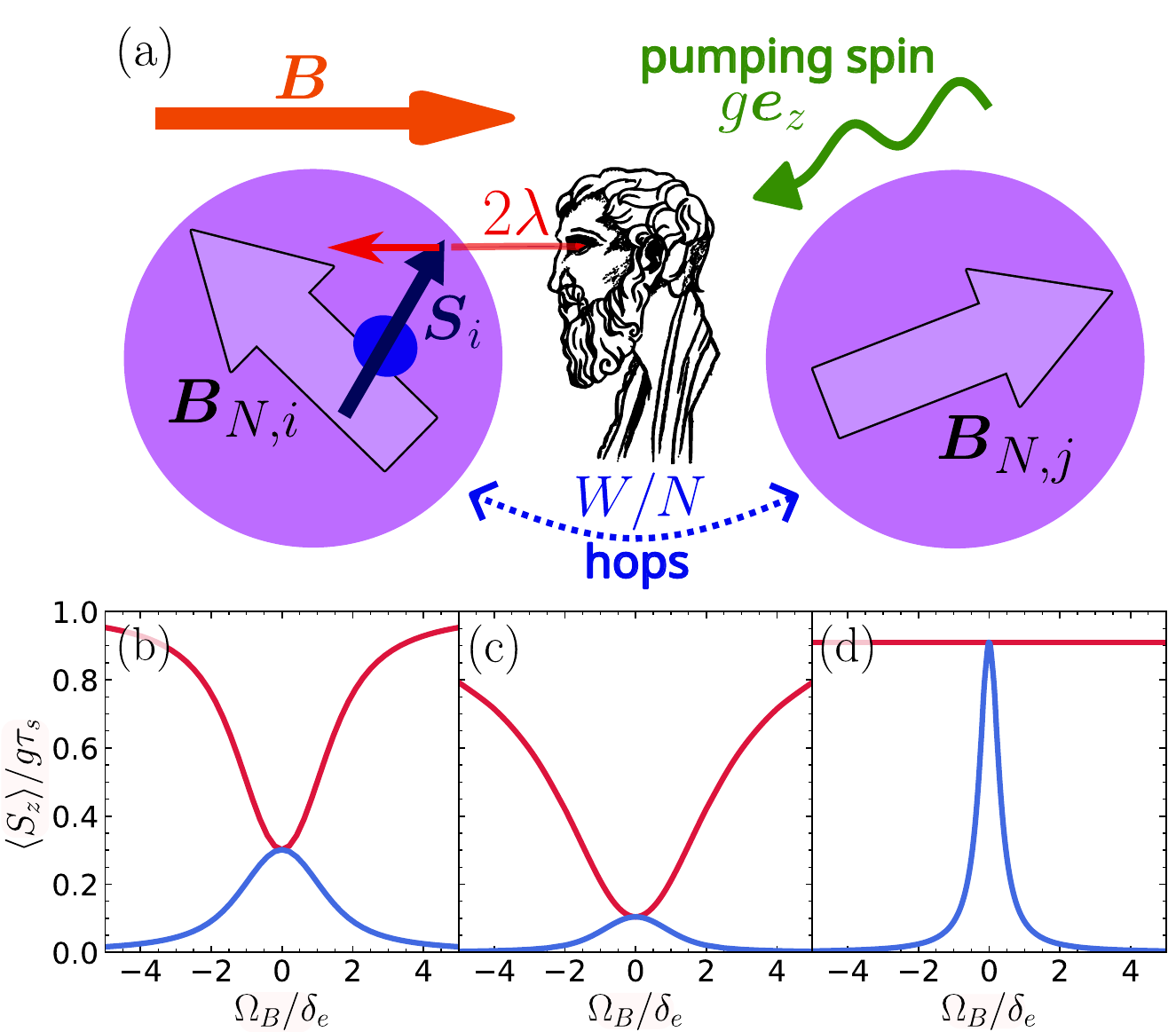}
		\end{center}
		\caption{\label{fig:model}
			(a) Sketch of the model: quantum dots (purple circles) under continuous optical spin pumping (green wavy arrow) and measurement (red arrow). Small blue circle with a black arrow represents an electron and its spin. Large contoured arrows denote the effective nuclear magnetic field. Blue dotted and orange arrows stand for electron hopping and external magnetic field, respectively. Three lower panels show polarization recovery (red) and Hanle (blue) curves calculated after Eqs.~\eqref{eq:SIZF} and~\eqref{eq:Voigt} for $\delta_e\tau_s=100$ and different hopping and measurement rates: $W=2\lambda=0$ (b); $W=0.01\delta_e$, $2\lambda = 0.05\delta_e$ (c); $W=1000\delta_e$, $2\lambda=10\delta_e$ (d).
		}
	\end{figure}

	We consider an ensemble of localized electrons under optical measurement of their spin dynamics. We take into account a few experimentally important process: (i) Larmor spin precession in external magnetic field~\cite{dyakonov_book}, (ii) hyperfine interaction with the spin fluctuations of the nuclei~\cite{GlazovBook2018}, (iii) electron hopping between localization sites~\cite{Efros89_eng}, (iv) optical electron spin pumping~\cite{OptOr}, (v) unavoidable measurement back action~\cite{khalfin1958contribution,doi:10.1063/1.523304}. They are shown schematically in Fig.~\ref{fig:model}(a), and each of them separately was already described microscopically. Together, they give rise to the following phenomenological system of kinetic equations for the spin dynamics of ensemble of localized electrons:
	\begin{multline}
		\label{eq:main_eq}
		\frac{\d \bm{S}_i}{\d t}=(\bm{\Omega}_B+\bm{\Omega}_{N,i})\times \bm{S}_i - \frac{\bm{S}_i}{\tau_s} \\ + W\left(\braket{\bm S}-\bm S_i\right) - 2\lambda \bm{S}_{i,\perp} + \bm{g}.
	\end{multline}
	Here $\bm S_i$ are the electron spins at the corresponding localization sites (quantum dots). $\bm\Omega_B=g\mu_B\bm B/\hbar$ is the Larmor frequency of electrons in external field $\bm B$ with $g$ being the electron effective $g$-factor and $\mu_B$ being the Bohr magneton. $\bm\Omega_i$ are the electron spin precession frequencies in the stochastic Overhauser field $\bm B_i$, produced by the nuclear spin fluctuations. The last term in the first line of Eq.~\eqref{eq:main_eq} describes the on-site electron spin relaxation unrelated to the hyperfine interaction~\cite{PhysRevB.64.125316,PhysRevB.66.161318}.
	
	The first term in the second line of Eq.~\eqref{eq:main_eq} describes electron hopping between $N$ sites with the rate $W$. It leads to the outgoing term and ingoing one with the average spin polarization of ensemble $\braket{\bm S}=\sum_i\bm S_i/N$~\cite{Glazov2015}. Further, $\lambda$ is the measurement strength, which describes the back action of both spin measurement and spin pumping with $\bm S_{i,\perp}=S_{i,x}\bm e_x+S_{i,y}\bm e_y$~\cite{Leppennen2022,PRC,Zeno_exp}. We denote the optical axis as $z$ and $\bm e_\alpha$ is the unit vector along the $\alpha$ axis ($\alpha=x,y,z$). The transverse spin components decay because they do not commute with the longitudinal spin $S_{i,z}$, which is measured~\cite{PhysRevA.36.5543,PRESILLA199695,PhysRevA.84.030101,PhysRevB.103.045413}. Finally, the last term in Eq.~\eqref{eq:main_eq} stands for the electron spin pumping with the rate $\bm g=g\bm e_z$ along the optical axis due to the optical orientation~\cite{Zhukov07,PhysRevB.98.125306}.
	
	The strength of the measurement back action $\lambda$ is given by the population of the trion state divided by its lifetime $\tau_0$~\cite{Leppennen2022}. Assuming resonant excitation and neglecting all inelastic processes except for the radiative recombination, this gives
	\begin{equation}
		\lambda=2\tau_0(M/\hbar)^2,
	\end{equation}
	where $M$ is the optical transition matrix element. It equals to the product of the amplitude of the electric field $E$ and the transition dipole moment $d$. The latter determines also the recombination rate as~\cite{yugova09}
	\begin{equation}
		\frac{1}{\tau_0}=\frac{4\sqrt{\eps}\omega_0^3d^2}{3\hbar c^3},
	\end{equation}
	where $\omega_0$ is the trion resonance frequency and $\eps$ is the high frequency dielectric constant. Thus, the dipole moment $d$ cancels in the expression for the measurement strength and we obtain a simple and universal expression
	\begin{equation}
		\lambda=\frac{3c^3E^2}{2\sqrt{\eps}\hbar\omega_0^3}.
	\end{equation}
	It is useful to express here the amplitude of the electric field through the incident light power $I$ as
	\begin{equation}
		E^2=\frac{8\pi I}{(1+\sqrt{\eps})^2c},
	\end{equation}
	where we took into account the Fresnel formula for the transmission coefficient of light through the sample surface. This gives the measurement strength
	\begin{equation}
		\label{eq:lambda_estimation}
		\lambda = \frac{12\pi}{\sqrt{\varepsilon}(1+\sqrt{\varepsilon})^2}\frac{c^2}{\hbar\omega_{0}^3}I.
	\end{equation}
	For the typical values of $\hbar \omega_{0} \sim 1$~eV and $\varepsilon \sim 10$ we obtain:
	\begin{equation}
		\frac{\lambda}{I} \sim \frac{\mathrm{~ns}^{-1}}{\mathrm{~W/cm}^{2}} .
	\end{equation}
	This shows that the back action can be quite strong for moderate light powers $I\sim 1\mathrm{~W/cm}^2$ already. Note that for larger intensities, the linear relation between $I$ and $\lambda$ breaks down, and the measurement strength saturates.
	
	Eqs.~\eqref{eq:main_eq} imply a number of assumptions. We specify them here and discuss their possible violation in Sec.~\ref{sec:Discussion}. Firstly, the ensemble is assumed to be homogeneous, that is the parameters of the spin dynamics are the same at each site. Further, we neglect the spatial correlations and memory effects for the hopping by assuming that the average electron spin contributes to the spin income to every site. The interaction between electrons is also neglected under the assumption of low electron concentration as compared to the number of available sites. The optical excitation is assumed to be (quasi-)resonant with the trion resonance. The trions are assumed to have a short lifetime, so that their population is negligible. Nevertheless, the trion (virtual) excitation leads to the decay of the transverse electron spin components according to the optical selection rules and in agreement with the phenomenology of the quantum measurement back action. Also we assume a small probability of the spin flip during the trion lifetime, which produces the electron spin polarization under excitation by circularly polarized light. Thus, both terms $\lambda$ and $g$ can be tuned by the light power.
	
	Moreover, we note that since the nuclear spin dynamics is much slower than the electron spin dynamics, $\bm\Omega_i$ can be considered static~\cite{Merkulov2002,GlazovBook2018}. They are normally distributed with the probability density function
	\begin{equation}
		\label{eq:nuclear_distribution}
		\mathcal{F}(\bm{\Omega})=\frac{1}{(\sqrt{\pi}\delta_e)^3}\exp{\left(-\frac{\Omega^2}{\delta_e^2}\right)},
	\end{equation}
	where parameter $\delta_e$ describes the dispersion.
	
	To solve Eqs.~\eqref{eq:main_eq} in a general case, it is convenient to rewrite it as
	\begin{equation}
		\label{eq:S_K}
		\frac{\d\bm S_i}{\d t}=\bm g+W\braket{\bm S}-\hat{\mathcal K_i}\bm S_i
	\end{equation}
	with the kinetic matrices
	\begin{equation}
		\mathcal K_{i,\alpha\beta}=\epsilon_{\alpha\beta\gamma}\Omega_{i,\gamma}+\delta_{\alpha\beta}\left[\left(\frac{1}{\tau_s}+W\right)+2\lambda\left(1-\delta_{\alpha z} \right)\right],
	\end{equation}
	where $\bm\Omega_i=\bm\Omega_{N,i}+\bm\Omega_B$, $\epsilon_{\alpha\beta\gamma}$ is the Levi-Civita symbol, and $\delta_{\alpha\beta}$ is the Kronecker delta. Now, the formal solution of Eqs.~\eqref{eq:S_K} gives the self consistency equation for the average spin:
	\begin{equation}
		\braket{\bm S}=\braket{\hat{\mathcal K}_i^{-1}}\left(\bm g+W\braket{\bm S}\right),
	\end{equation}
	where the averaging is performed with the distribution function~\eqref{eq:nuclear_distribution}. Its solution has the form
	\begin{equation}
		\label{eq:S_general}
		\braket{\bm S}=\frac{\braket{\hat{\mathcal K}_i^{-1}}\bm g}{1-\braket{\hat{\mathcal K}_i^{-1}}W}.
	\end{equation}
	The experimentally relevant observable in this setup is $\braket{S_z}$, we analyze it in detail in the rest of the paper.

	\section{Results}
	\label{sec:Results}
	
	\subsection{Zero and longitudinal magnetic field}
	\label{subsec:SIZF}
	
	We start by considering the Faraday geometry, when external magnetic field is parallel to the optical axis $z$ (or absent) and the system is axially symmetric. In this case, the average spin is directed along the $z$ axis and one has to calculate $\braket{\hat{\mathcal K}^{-1}_i}_{zz}$ only. Then the average spin is given by
	\begin{equation}
		\label{eq:SIZF}
		\frac{\braket{S_z}}{g\tau_s}=\frac{\mathcal{A}}{W_0\tau_s(1-\mathcal{A})+1},
	\end{equation}
	where $\mathcal A=\left(W+1/\tau_s\right)\langle\hat{\mathcal K}^{-1}_i\rangle_{zz}$. Explicitly,
	\begin{equation}
		\label{eq:A}
		\mathcal{A}=\left\langle \frac{1 + T_{\perp}^2\Omega_{i,z}^2}{1 + T_zT_{\perp}\Omega_{i,\perp}^2 + T_{\perp}^2\Omega_{i,z}^2} \right\rangle,
	\end{equation}
	where $T_z^{-1}=W+\tau_s^{-1}$ and $T_{\perp}^{-1}=W+2\lambda+\tau_s^{-1}$ denote the effective longitudinal and transverse spin lifetimes at the given site and $\Omega_{i,\perp}^2=\Omega_{i,x}^2+\Omega_{i,y}^2$. In particular cases of zero back action ($\lambda=0$) and absence of hopping ($W=0$), Eq.~\eqref{eq:SIZF} agrees with the results obtained in Refs.~\onlinecite{PRC} and~\onlinecite{Leppennen2022}, respectively.
	
	For example, if the spin relaxation is unrelated to the nuclei, $\tau_s\delta\ll 1$, we obtain $\mathcal A=1$ and Eq.~\eqref{eq:SIZF} gives $\braket{S_z}=g\tau_s$, as expected. Moreover, the same happens in the limit of a strong magnetic field, when $\Omega_B$ is large. Namely, $\mathcal A$ goes again to one, and the spin polarization equals $g\tau_s$. This effect of suppression of the nuclei-induced spin relaxation by the longitudinal magnetic field termed polarization recovery~\cite{PRC}.
	
	The examples of polarization recovery curves are shown in Fig.~\ref{fig:model}(b---d) by the red lines for the typical case of large product $\delta_e\tau_s$. One can see that depending on the hopping rate and measurement back strength, both the amplitude and the width of the polarization recovery curve change significantly.
	
	In the basic limit, when $W=\lambda=0$ and electron spin relaxation is driven by the hyperfine interaction only~\cite{Merkulov2002}, the polarization recovery curve is described by~\cite{petrov08,eh_noise}
	\begin{equation}
		\label{eq:PRC_Mer}
		\frac{\braket{S_z}}{g\tau_s}\approx\frac{1}{3}\frac{2\delta_e^2+3\Omega_B^2}{2\delta_e^2+\Omega_B^2}.
	\end{equation}
	So it has a Lorentzian shape with the width determined by the typical nuclear field fluctuation $\delta_e$. In this case, the electron spin relaxation is caused by the precession in a static and random nuclear field. In zero field, one has $\braket{S_z}=g\tau_s/3$, which corresponds to the conservation of the one third of the electron spin polarization being parallel to the random spin precession direction $\bm\Omega_{N,i}$.
	
\begin{figure*}[htpb]
		\begin{center}
			\includegraphics[width=0.7\linewidth]{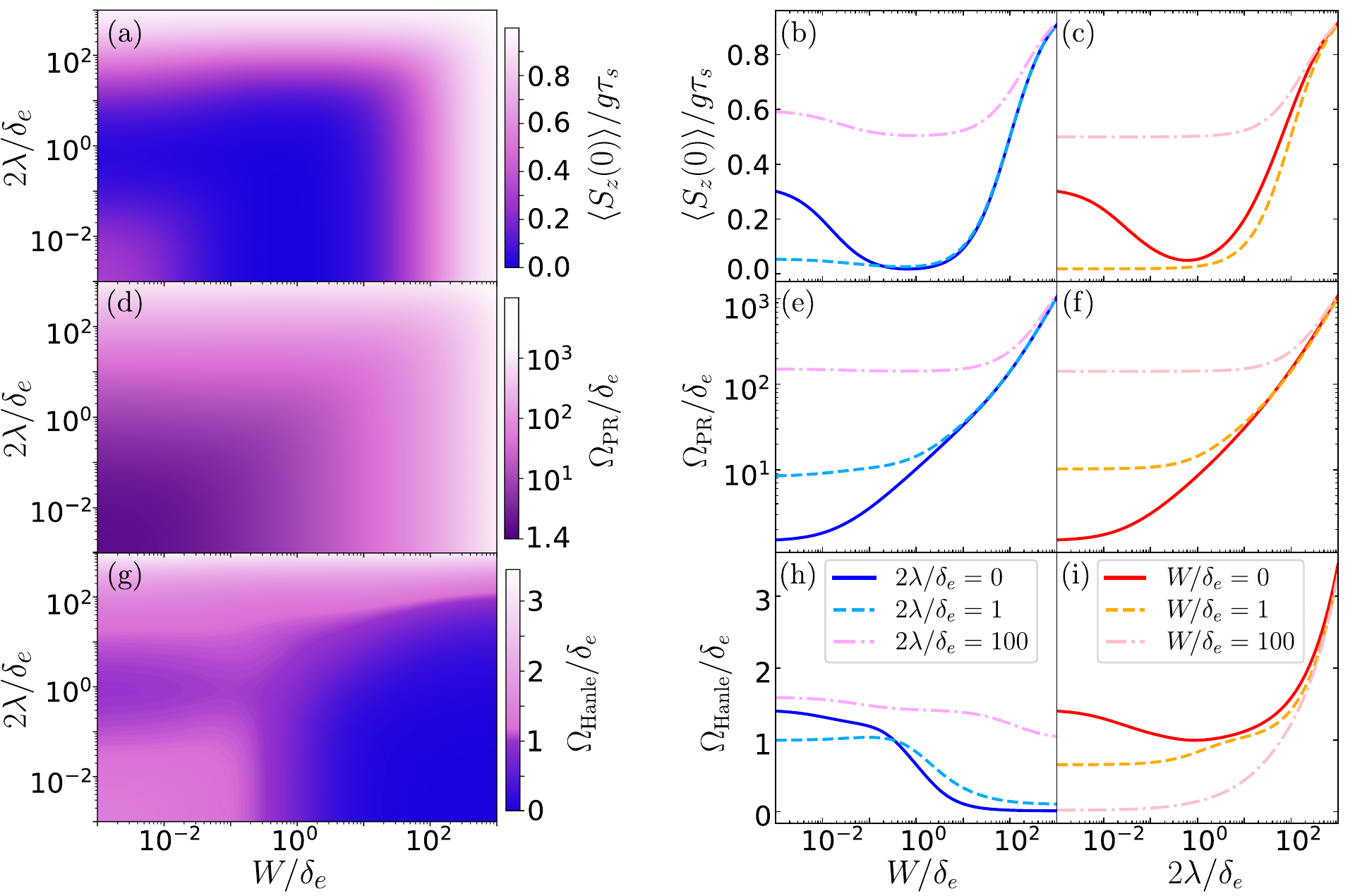}
		\end{center}
		\caption{\label{fig:results_union}
			Dependence of the parameters of Hanle and polarization recovery curves on the hopping rate $W$ and measurement back action strength $\lambda$ (calculated for $\delta_e\tau_s=100$). (a---c) Show the depth of the polarization recovery curve, which is also the amplitude of the Hanle curve. (d---f) and (g---i) show HWHM of polarization recovery and Hanle curves, respectively. The left column, (a), (d), (g), shows dependences on both $W$ and $\lambda$ as the color maps. The middle column, (b), (e), (h) and the right column, (c), (f), (i), show the dependences on $W$ and $\lambda$ for a few values $\lambda$ and $W$, respectively, as indicated in the legend.
		}
	\end{figure*}

	The dependence of the depth of the polarization recovery curve on the hopping rate and measurement strength is shown in Fig.~\ref{fig:results_union}(a---c). Generally, they are qualitatively similar. The spin polarization in zero field, $\braket{S_z(0)}$, nonmonotonously depends on the measurement strength $\lambda$ in agreement with the general phenomenology of the quantum back action~\cite{Leppennen2022}: The weak measurements lead to the relaxation of the electron spin component parallel to the local precession frequency $\bm\Omega_{N,i}$. This accelerates the spin relaxation and leads to the decrease of the electron spin polarization, which is the quantum anti-Zeno effect. By contrast, the strong measurements continuously project electron spin to the optical axis, which suppresses electron spin precession and dephasing. This is the quantum Zeno effect. As a result, the electron spin polarization increases and tends to $g\tau_s$ in the limit $\lambda\to\infty$, as one can see in the figures.
	
	The electron hopping has a similar effect. Slow hopping leads to the rare changes of the nuclear field experienced by the given electron. This breaks the conservation of one third of the spin polarization and accelerates spin relaxation rate by $2W/3$~\cite{GlazovBook2018,PhysRevLett.134.016201}. The spin polarization reaches minimum at $W\sim\delta_e$ and further increase of the hopping rate leads to the motional narrowing regime (similar to the Dyakonov-Perel spin relaxation~\cite{DyakonovPerel1971}), when the spin relaxation slows down with increase of $W$. Generally, when the condition $W(W + 2\lambda) \gg \delta_e^2$ is satisfied, we find from Eq.~\eqref{eq:SIZF} that the spin polarization can be written as
	\begin{equation}
		\label{eq:SIZF_hopping_limit}
		S_z=gT_1
	\end{equation}
	with the effective longitudinal spin relaxation rate
	\begin{equation}
		\label{eq:T_1}
		\frac{1}{T_1} = \frac{1}{\tau_s} + \frac{\delta_e^2}{W+2\lambda}.
	\end{equation}
	One can say, that the measurement back action effectively renormalizes the hopping rate in this regime.
	
	Fig.~\ref{fig:results_union}(d---f) shows the dependence of the polarization recovery curve half width at half maximum (HWHM), $\Omega_{\text{PR}}$, on the hopping rate and the measurement strength. The larger both of them, the broader the polarization recovery curve. Indeed, at small $\lambda$ and $W$, when the measurement back action and hopping accelerate spin relaxation, stronger magnetic field is needed to stabilize the electron spin along the $z$ axis. Also at large $\lambda$ and $W$, in the motional narrowing regime, the Larmor precession frequency should be larger than the correlation time to suppress the spin relaxation. The analysis of Eq.~\eqref{eq:SIZF} reveals that
	\begin{equation}
		\label{eq:Omega_PR_hopping_limit}
		\Omega_{\text{PR}} = W+2\lambda,
	\end{equation}
	which shows again that $W$ and $\lambda$ have the same effect.
	
	To conclude this subsection, measurement back action and hopping qualitatively sum up in the dynamics of the electron spin in the longitudinal magnetic field.

	\subsection{Voigt geometry}
	\label{subsec:Voigt_geometry}
	
	In transverse magnetic field, we choose the $x$ axis to be parallel to $\bm B$, so that the average spin lies in the $(yz)$ plane. Thus, one has to calculate the components $\braket{\hat{\mathcal K}^{-1}_{\alpha\beta}}$ with $\alpha,\beta=y,z$. Omitting the cumbersome intermediate expressions, Eq.~\eqref{eq:S_general} gives
	\begin{widetext}
		\begin{equation}
			\label{eq:Voigt}
			\frac{\braket{S_z}}{g\tau_s} = \frac{\mathcal{A}[1+W_0\tau_s(1-\mathcal{A}+\mathcal{D})]-W_0\tau_s\mathcal{B}^2}{[1+W_0\tau_s(1-\mathcal{A}+\mathcal{D})][1+W_0\tau_s(1-\mathcal{A})]-W_0^2\tau_s^2\mathcal{B}^2}
		\end{equation}
	\end{widetext}
	where $\mathcal{B}$ and $\mathcal{D}$ describe the roles of external magnetic field and measurement-induced anisotropy of the spin relaxation, respectively. Explicitly, they are given by
	\begin{subequations}\label{eq:B_and_D}
		\begin{align}
			\mathcal{B} &=\left\langle \frac{T_{\bot} \Omega_{i,x}}{1 + T_zT_{\bot}\Omega_{i,\bot}^2 + T_{\bot}^2 \Omega_{i,z}^2} \right\rangle, \label{eq:B} \\[5pt]
			\mathcal{D} &=\left\langle \frac{1-T_z/T_{\bot}}{1 + T_zT_{\bot}\Omega_{i,\bot}^2 + T_{\bot}^2 \Omega_{i,z}^2} \right\rangle \label{eq:D}.
		\end{align}
	\end{subequations}
	We note, that in the limit of zero measurement strength, this result agrees with Ref.~\onlinecite{PRC}.
	
	The transverse magnetic field suppresses electron spin polarization, which is known as the Hanle effect~\cite{OptOr,dyakonov_book}. The examples of the Hanle curves are shown by the blue lines in Fig.~\ref{fig:model}(b---d). For instance, when measurement back action and hopping are neglected, $\lambda=W=0$, one has~\cite{petrov08,eh_noise}
	\begin{equation}
		\frac{\braket{S_z}}{g\tau_s}=\frac{2}{3}\frac{\delta_e^2}{2\delta_e^2+\Omega_B^2}.
	\end{equation}
	So the Hanle curve is Lorentzian with the HWHM $\Omega_{\text{Hanle}}$ equal to $\sqrt{2}\delta_e$. Naturally, it corresponds to the electron spin dephasing time of the order of $\delta_e^{-1}$.
	
	Fig.~\ref{fig:results_union}(g---i) shows the dependence of $\Omega_{\text{Hanle}}$ on the hopping rate and measurement strength. One can see that generally it is quite complex. Let us describe it in detail.
	
	In the absence of the back action, the increase of the electron hopping leads to the monotonous decrease of the width of the Hanle curve, see dark blue curve in Fig.~\ref{fig:results_union}(h). In fact, when $W\ll\delta_e$ it has hardly any effect, because the spin precession period is much smaller than the correlation time. At $W\gtrsim\delta_e$, the hopping slows down the electron spin precession and, accordingly, the width of the Hanle curve decreases.
	
	The measurement back action has a more complicated effect. On one hand, it leads to the nonmonotoneous change of the longitudinal spin relaxation time, due to the quantum anti-Zeno and Zeno effects discussed in the previous subsection. On the other hand, increase of $\lambda$ leads to the monotonous acceleration of the transverse spin relaxation due to the term $2\lambda$ in Eq.~\eqref{eq:main_eq}. The interplay of these two effects results in the nonmonotoneous change of the width of the Hanle curve with minimum in the regime of the anti-Zeno effect.
	
	This situation changes for the fast electron hopping, $W>\delta_e$. In this case, only Zeno effect takes place, because the electron spin does not recover between hops. The measurements push spin to the $z$ axis only, which results in the broadening of the Hanle curve. From the analysis of Eq.~\eqref{eq:Voigt} in the corresponding limit $\left(W(W+2\lambda) \gg \delta_e\right)$, we obtain
	\begin{equation}
		\label{eq:Omega_Hanle}
		\Omega_{\text{Hanle}}=\sqrt{\frac{1}{T_1 T_2}}
	\end{equation}
	with the effective transverse spin relaxation rate
	\begin{equation}
		\label{eq:T_2}
		\frac{1}{T_2} = \frac{1}{\tau_s} + \frac{\delta_e^2}{W} + 2\lambda.
	\end{equation}
	Eq.~\eqref{eq:Omega_Hanle} is usual for the width of the Hanle curve in the case of the anisotropic spin relaxation with the longitudinal time $T_1$ and transverse time $T_2$. The latter differs from Eq.~\eqref{eq:T_1} by $2\lambda$ as expected for the measurement back action, see Eq.~\eqref{eq:main_eq}.
	
	Thus, from Fig.~\ref{fig:results_union}(g) and comparison of panels (h) and (i), we conclude that in the transverse magnetic field, the measurement back action and hopping compete with each other in contrast to the Faraday geometry.

	\section{Discussion} 
	\label{sec:Discussion}
	
\begin{figure*}[htpb]
		\begin{center}
			\includegraphics[width=1.0\linewidth]{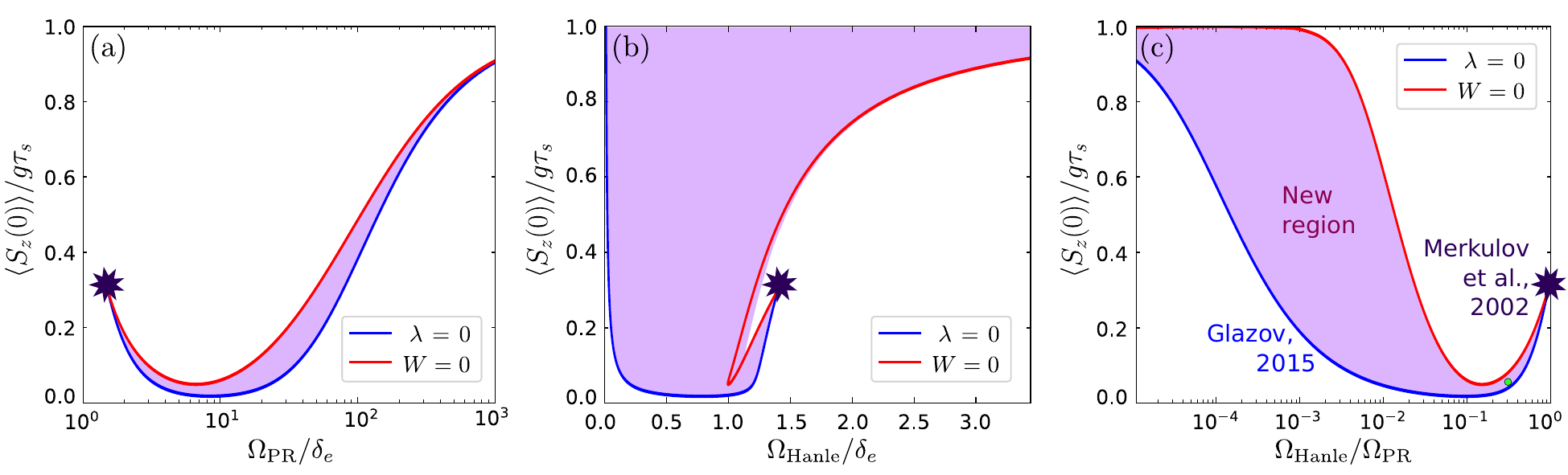}
		\end{center}
		\caption{\label{fig:discussion_plots}
			Regions of the parameters of Hanle and polarization recovery curves, which are possible in the model: $\braket{S_z(0)}$ and $\Omega_{\text{PR}}$ (a), $\braket{S_z(0)}$ and $\Omega_{\text{Hanle}}$ (b), $\braket{S_z(0)}$ and $\Omega_{\text{Hanle}}/\Omega_{\text{PR}}$ (c). The blue and red lines show the parameters for $\lambda=0$ and $W=0$, respectively. A small green dot in (c) shows the parameters of Hanle and polarization recovery curves from Ref.~\onlinecite{kirsteinExtendedSpinCoherence2021}.}
	\end{figure*}

	Let us discuss the practical outcome of the obtained results. Typically, the exact parameters of the spin dynamics are not known, which means that they should be obtained from the fits of the polarization recovery and Hanle curves. Fig.~\ref{fig:discussion_plots}(a) shows the ranges of possible values of polarization recovery curve depth and width. The star denotes the basic limit of zero hopping and back action, Eq.~\eqref{eq:PRC_Mer}. The blue curve corresponds to the case when only hopping is taken into account~\cite{Glazov2015}. The complementary limit of back action only without hopping is shown by the red curve. These two curves are quite close to each other, because hopping and back action have a similar effect in the longitudinal magnetic field. Accordingly, the range of the possible value of $\braket{S_z}$ and $\Omega_{\text{PR}}$ between them is quite narrow.
	
	In a similar way, Fig.~\ref{fig:discussion_plots}(b) shows the possible range of the Hanle curve amplitude and width. The notations here are the same as in the panel (a). However, due to the qualitatively different dependence of the Hanle curve width on the hopping rate $W$ and measurement strength $\lambda$, the blue and red curves are no longer close to each other. Accordingly, the range of possible widths and amplitudes of the Hanle curves is broad. Interestingly, it is not limited by the blue and red curves, because the map $(W,\lambda)\mapsto(\Omega_{\text{Hanle}},\braket{S_z})$ is not a diffeomorphism.
	
	Usually, the strength of the hyperfine interaction is not known exactly. For this reason, we show in Fig.~\ref{fig:discussion_plots}(c) the range of possible amplitude of the spin dependences and the ratio of the Hanle and polarization recovery curves widths. This range is also quite broad. We stress, that in the previous models, these two parameters were strictly related, as, for example, for the blue curve. Therefore a possibility to describe the experimental results was quite limited. The present theory significantly broadens the range of experimental data, which can be described theoretically, if both electron hopping and measurement back action are taken into account.
	
	The developed theory can be applied to the description of the experiments with various systems including, self-assembled quantum dots~\cite{PhysRevLett.94.116601,petrov08}, shallow donors~\cite{PhysRevLett.88.256801,kirsteinExtendedSpinCoherence2021}, localized states in the quantum wells produced by the interface imperfections~\cite{PRC,noise-trions}, and some organic molecules~\cite{shumilinMicroscopicTheoryOrganic2020,noise_OMAR}. Specifically, we find that it might be relevant for the experiments performed in Refs.~\onlinecite{mikhailovElectronHoleSpin2018,kudlacikOpticalSpinOrientation2024,auerMeasurementKnightField2009,PhysRevLett.88.256801,dzhioevStabilizationElectronNuclearSpin2007,heisterkampLongitudinalTransverseSpin2015,kirsteinExtendedSpinCoherence2021}. For example, the parameters of the curves from Ref.~\onlinecite{kirsteinExtendedSpinCoherence2021} are $\Omega_{\mathrm{Hanle}}/\Omega_{\mathrm{PR}}\approx 0.25$ and $S_z(0)/g\tau_s\approx0.05$. They are shown by a green dot in Fig.~\ref{fig:discussion_plots}(c). One can see, that they are inconsistent with the models, which takes into account only hopping and only back action. However, for the reliable interpretation of the experiments, a more detailed analysis and measurements for the different light powers are needed.
	
	The proposed theoretical model is so simple that is allows for analytical solution and analysis. However, it does not take into account many aspects of realistic systems. For example, the inhomogeneity of the system is neglected. In a more realistic model, the parameters $\lambda,g,\tau_s,\delta_e$ can be different for different sites. Nevertheless, they can be expected to be correlated with the electron localization energy~\cite{yugova09}. The electron hopping is most efficient between sites with a similar energy and, thus, similar parameters. Additionally, under the conditions of the resonant optical spin measurement, only a small quasi homogeneous subensemble of resonant sites is addressed. Therefore, the hops out of this subensemble can be described by the renormalization of the isotropic spin relaxation time $\tau_s$. We note also that to some extent, hopping can effectively describe reorientation of the nuclear spin fluctuations due to the quadrupole splittings, the Knight field, or the nuclear spin-spin interactions.
	
	In the description of the electron hopping, the electron spin rotations during the hops are neglected. This is a realistic assumption, because the corresponding rotation angles are typically small due to the weakness of the spin-orbit coupling~\cite{KKavokin-review,lyubinskiy07,Hopping_spin}. However, they can become important in the limit of fast hopping, when the nuclei-induced spin relaxation gets suppressed. Also, the memory effects are neglected for the hopping. In practice, the distribution of the hopping rates is typically exponentially broad being proportional to the tunneling exponent. However, this effect is not expected to be important, when the hopping rate is small, $W\ll\delta_e$, because a few hops lead already to complete spin relaxation~\cite{Shumilin2015,PhysRevB.100.075409}. Similarly, in the opposite limit of $W\gg\delta_e$, the return probability does not play a role, because an electron visits many sites during its spin relaxation time.
	
	An interesting feature which is not analyzed in this paper is the slightly non-Lorentzian shape of the polarization recovery and Hanle curves. This can be directly checked for the experimental measurements. However, we expect that the inhomogeneity of the ensemble can significantly contribute to the distortion of the shape also.
	
	Finally, we note that the measurement back action being unavoidable under the optical pumping and probing the electron spin ensemble leads to the additional relaxation of the transverse spin components. In principle, there can be other sources of the anisotropy of the electron spin relaxation. They can be included in the renormalization of the measurement strength $\lambda$, but they are less easily controlled in the experiments.
	
	As an outlook, our model of the spin dynamics can be solved similarly in a non-stationary case using the Fourier transform. This gives a possibility to apply it for the description of the pump-probe experiments~\cite{glazov:review} and the measurements of the electron spin noise spectra~\cite{UspFizNauk2021}.

	\section{Conclusion}
	\label{sec:Conclusion}
	
	We have developed a model of the polarization recovery and Hanle effects for ensembles of localized electrons taking into account the interplay between electron hopping and quantum measurement back action. We find that they have a similar effect on the spin dynamics in the longitudinal magnetic field and effectively sum up. By contrast, they compete in the transverse magnetic field, which results in the parametrically broad family of experimental curves, which can be described by the developed model. Moreover, the same model can be applied for the description of the non-stationary electron spin dynamics and the spin noise.

	\acknowledgements
	
	We are deeply grateful to D. Zibinskaya for the drawing of Zeno's profile. We thank the Foundation for the Advancement of Theoretical Physics and Mathematics “BASIS.” The work is supported by the Russian Science Foundation Grant No. 25-72-10031. 
	
	\renewcommand{\i}{\ifr}
	\bibliography{Zeno_hops_lib}

@article{auerMeasurementKnightField2009,
  title = {Measurement of the {{Knight}} Field and Local Nuclear Dipole-Dipole Field in an {{InGaAs}}/{{GaAs}} Quantum Dot Ensemble},
  author = {Auer, T. and Oulton, R. and Bauschulte, A. and Yakovlev, D. R. and Bayer, M. and Verbin, S. {\relax Yu}. and Cherbunin, R. V. and Reuter, D. and Wieck, A. D.},
  year = {2009},
  month = nov,
  journal = {Phys. Rev. B},
  volume = {80},
  number = {20},
  pages = {205303},
  publisher = {American Physical Society},
  doi = {10.1103/PhysRevB.80.205303},
  urldate = {2025-11-10}
}

@article{BAP,
  title = {Spin Relaxation of Electrons Due to Scattering by Holes},
  author = {Bir, G. L. and Aronov, A. G. and Pikus, G. E.},
  year = {1975},
  journal = {Sov. Phys. JETP},
  volume = {42},
  number = {4},
  pages = {705},
  publisher = {AIP}
}

@article{berezovsky2006nondestructive,
  title = {Nondestructive Optical Measurements of a Single Electron Spin in a Quantum Dot},
  author = {Berezovsky, J. and Mikkelsen, M. H. and Gywat, O. and Stoltz, N. G. and Coldren, L. A. and Awschalom, D. D.},
  year = {2006},
  journal = {Science},
  volume = {314},
  number = {5807},
  pages = {1916},
  publisher = {American Association for the Advancement of Science}
}

@article{burkardSemiconductorSpinQubits2023,
  title = {Semiconductor Spin Qubits},
  author = {Burkard, Guido and Ladd, Thaddeus D. and Pan, Andrew and Nichol, John M. and Petta, Jason R.},
  year = {2023},
  month = jun,
  journal = {Rev. Mod. Phys.},
  volume = {95},
  number = {2},
  pages = {025003},
  publisher = {American Physical Society},
  doi = {10.1103/RevModPhys.95.025003},
  urldate = {2025-11-05}
}

@article{bussElectronSpinDynamics2016,
  title = {Electron Spin Dynamics in Cubic {{GaN}}},
  author = {Bu{\ss}, J. H. and Schupp, T. and As, D. J. and Brandt, O. and H{\"a}gele, D. and Rudolph, J.},
  year = {2016},
  month = dec,
  journal = {Phys. Rev. B},
  volume = {94},
  number = {23},
  pages = {235202},
  publisher = {American Physical Society},
  doi = {10.1103/PhysRevB.94.235202},
  urldate = {2025-11-10}
}

@article{doi:10.1063/1.523304,
  title = {The {{Zeno}}'s Paradox in Quantum Theory},
  author = {Misra, B. and Sudarshan, E. C. G.},
  year = {1977},
  journal = {J. Math. Phys.},
  volume = {18},
  number = {4},
  pages = {756},
  doi = {10.1063/1.523304}
}

@article{DP,
  title = {Spin Orientation of Electrons Associated with Interband Absorption of Light in Semiconductors},
  author = {Dyakonov, M.I. and Perel, V.I.},
  year = {1971},
  journal = {Sov. Phys. JETP},
  volume = {33},
  number = {5},
  pages = {1053}
}

@book{dyakonov_book,
  title = {Spin Physics in Semiconductors},
  editor = {Dyakonov, M. I.},
  year = {2017},
  publisher = {Springer International Publishing AG, Berlin}
}

@article{dyakonov72,
  title = {Spin Relaxation of Conduction Electrons in Noncentrosymmetric Semiconductors},
  author = {Dyakonov, M.I. and Perel', V.I},
  year = {1972},
  journal = {Sov. Phys. Solid State},
  volume = {13},
  pages = {3023}
}

@article{DyakonovPerel1971,
  title = {Spin Relaxation of Conduction Electrons in Noncentrosymmetric Semiconductors},
  author = {Dyakonov, M. I. and Perel, V. I.},
  year = {1971},
  journal = {Sov. Phys. Solid State},
  volume = {13},
  pages = {3581}
}

@article{dzhioevStabilizationElectronNuclearSpin2007,
  title = {Stabilization of the {{Electron-Nuclear Spin Orientation}} in {{Quantum Dots}} by the {{Nuclear Quadrupole Interaction}}},
  author = {Dzhioev, R. I. and Korenev, V. L.},
  year = {2007},
  month = jul,
  journal = {Phys. Rev. Lett.},
  volume = {99},
  number = {3},
  pages = {037401},
  publisher = {American Physical Society},
  doi = {10.1103/PhysRevLett.99.037401},
  urldate = {2025-11-10}
}

@book{Efros89_eng,
  title = {Electronic Properties of Doped Semiconductors},
  author = {Efros, A. L. and Shklovskii, B. I.},
  year = {2013},
  publisher = {Springer Science \& Business Media, Berlin}
}

@article{eh_noise,
  title = {Spin Noise of Electrons and Holes in ({{In}},{{Ga}}){{As}} Quantum Dots: {{Experiment}} and Theory},
  author = {Glasenapp, {\relax Ph}. and Smirnov, D. S. and Greilich, A. and Hackmann, J. and Glazov, M. M. and Anders, F. B. and Bayer, M.},
  year = {2016},
  month = may,
  journal = {Phys. Rev. B},
  volume = {93},
  number = {20},
  pages = {205429},
  publisher = {American Physical Society},
  doi = {10.1103/PhysRevB.93.205429}
}

@article{Elliott54,
  title = {Theory of the Effect of Spin-Orbit Coupling on Magnetic Resonance in Some Semiconductors},
  author = {Elliott, R. J.},
  year = {1954},
  month = oct,
  journal = {Phys. Rev.},
  volume = {96},
  number = {2},
  pages = {266},
  publisher = {American Physical Society},
  doi = {10.1103/PhysRev.96.266}
}

@article{Facchi_2008,
  title = {Quantum {{Zeno}} Dynamics: Mathematical and Physical Aspects},
  author = {Facchi, P. and Pascazio, S.},
  year = {2008},
  month = oct,
  journal = {J. Phys. A: Math. Theor.},
  volume = {41},
  number = {49},
  pages = {493001},
  publisher = {IOP Publishing},
  doi = {10.1088/1751-8113/41/49/493001}
}

@article{gerardot08,
  title = {Optical Pumping of a Single Hole Spin in a Quantum Dot},
  author = {Gerardot, B. D. and Brunner, D. and Dalgarno, P. A and Ohberg, P. and Seidl, S. and Kroner, M. and Karrai, K. and G. Stoltz, N. and Petroff, P. M. and Warburton, R. J.},
  year = {2008},
  journal = {Nature},
  volume = {451},
  pages = {441}
}

@article{Glazov_hopping,
  title = {Spin Noise of Localized Electrons: {{Interplay}} of Hopping and Hyperfine Interaction},
  author = {Glazov, M. M.},
  year = {2015},
  month = may,
  journal = {Phys. Rev. B},
  volume = {91},
  number = {19},
  pages = {195301},
  publisher = {American Physical Society},
  doi = {10.1103/PhysRevB.91.195301}
}

@article{glazov:review,
  title = {Coherent Spin Dynamics of Electrons and Excitons in Nanostructures (a Review)},
  author = {Glazov, M. M.},
  year = {2012},
  journal = {Phys. Solid State},
  volume = {54},
  pages = {1}
}

@article{Glazov2015,
  title = {Spin Noise of Localized Electrons: {{Interplay}} of Hopping and Hyperfine Interaction},
  author = {Glazov, M. M.},
  year = {2015},
  month = may,
  journal = {Phys. Rev. B},
  volume = {91},
  number = {19},
  pages = {195301},
  publisher = {American Physical Society},
  doi = {10.1103/PhysRevB.91.195301}
}

@book{GlazovBook2018,
  title = {Electron \& Nuclear Spin Dynamics in Semiconductor Nanostructures},
  author = {Glazov, M. M.},
  year = {2018},
  publisher = {Oxford University Press},
  doi = {10.1093/oso/9780198807308.001.0001},
  isbn = {978-0-19-184509-3}
}

@article{hanleUeberMagnetischeBeeinflussung1924,
  title = {{{\"U}ber magnetische Beeinflussung der Polarisation der Resonanzfluoreszenz}},
  author = {Hanle, Wilhelm},
  year = {1924},
  month = dec,
  journal = {Z. Physik},
  volume = {30},
  number = {1},
  pages = {93--105},
  issn = {0044-3328},
  doi = {10.1007/BF01331827},
  urldate = {2025-11-05},
  langid = {ngerman}
}

@article{heisterkampLongitudinalTransverseSpin2015,
  title = {Longitudinal and Transverse Spin Dynamics of Donor-Bound Electrons in Fluorine-Doped {{ZnSe}}: {{Spin}} Inertia versus {{Hanle}} Effect},
  shorttitle = {Longitudinal and Transverse Spin Dynamics of Donor-Bound Electrons in Fluorine-Doped {{ZnSe}}},
  author = {Heisterkamp, F. and Zhukov, E. A. and Greilich, A. and Yakovlev, D. R. and Korenev, V. L. and Pawlis, A. and Bayer, M.},
  year = {2015},
  month = jun,
  journal = {Phys. Rev. B},
  volume = {91},
  number = {23},
  pages = {235432},
  publisher = {American Physical Society},
  doi = {10.1103/PhysRevB.91.235432},
  urldate = {2025-11-10}
}

@article{Hopping_spin,
  title = {Electrical Spin Orientation, Spin-Galvanic, and Spin-Hall Effects in Disordered Two-Dimensional Systems},
  author = {Smirnov, D. S. and Golub, L. E.},
  year = {2017},
  month = mar,
  journal = {Phys. Rev. Lett.},
  volume = {118},
  number = {11},
  pages = {116801},
  publisher = {American Physical Society},
  doi = {10.1103/PhysRevLett.118.116801}
}

@article{kenEffectElectricCurrent2020,
  title = {Effect of Electric Current on the Optical Orientation of Interface Electrons in {{AlGaAs}}/{{GaAs}} Heterostructures},
  author = {Ken, O. S. and Zhukov, E. A. and Akimov, I. A. and Korenev, V. L. and Kopteva, N. E. and Kalitukha, I. V. and Sapega, V. F. and Wieck, A. D. and Ludwig, A. and Schott, R. and Kusrayev, {\relax Yu}. G. and Yakovlev, D. R. and Bayer, M.},
  year = {2020},
  month = jul,
  journal = {Phys. Rev. B},
  volume = {102},
  number = {4},
  pages = {045302},
  publisher = {American Physical Society},
  doi = {10.1103/PhysRevB.102.045302},
  urldate = {2025-11-10}
}

@article{khalfin1958contribution,
  title = {Contribution to the Decay Theory of a Quasi-Stationary State},
  author = {Khalfin, L. A.},
  year = {1958},
  journal = {Sov. Phys. JETP},
  volume = {6},
  pages = {1053}
}

@article{kirsteinExtendedSpinCoherence2021,
  title = {Extended Spin Coherence of the Zinc-Vacancy Centers in {{ZnSe}} with Fast Optical Access},
  author = {Kirstein, Erik and Zhukov, Evgeny A. and Smirnov, Dmitry S. and Nedelea, Vitalie and Greve, Phillip and Kalitukha, Ina V. and Sapega, Viktor F. and Pawlis, Alexander and Yakovlev, Dmitri R. and Bayer, Manfred and Greilich, Alex},
  year = {2021},
  month = sep,
  journal = {Commun. Mater.},
  volume = {2},
  number = {1},
  pages = {91},
  publisher = {Nature Publishing Group},
  issn = {2662-4443},
  doi = {10.1038/s43246-021-00198-z},
  urldate = {2025-10-02},
  copyright = {2021 The Author(s)},
  langid = {english}
}

@article{KKavokin-review,
  title = {Spin Relaxation of Localized Electrons in N-Type Semiconductors},
  author = {Kavokin, K. V.},
  year = {2008},
  journal = {Semicond. Sci. Technol.},
  volume = {23},
  number = {11},
  pages = {114009}
}

@article{koptevaSpinDephasingElectrons2019,
  title = {Spin Dephasing of Electrons and Holes in Isotopically Purified {{ZnSe}}/({{Zn}},{{Mg}}){{Se}} Quantum Wells},
  author = {Kopteva, N. E. and Kirstein, E. and Zhukov, E. A. and Hussain, M. and Bhatti, A. S. and Pawlis, A. and Yakovlev, D. R. and Bayer, M. and Greilich, A.},
  year = {2019},
  month = nov,
  journal = {Phys. Rev. B},
  volume = {100},
  number = {20},
  pages = {205415},
  publisher = {American Physical Society},
  doi = {10.1103/PhysRevB.100.205415},
  urldate = {2025-11-10}
}

@article{kudlacikOpticalSpinOrientation2024,
  title = {Optical {{Spin Orientation}} of {{Localized Electrons}} and {{Holes Interacting}} with {{Nuclei}} in a {{FA0}}.{{9Cs0}}.{{1PbI2}}.{{8Br0}}.2 {{Perovskite Crystal}}},
  author = {Kudlacik, Dennis and Kopteva, Nataliia E. and Kotur, Mladen and Yakovlev, Dmitri R. and Kavokin, Kirill V. and Harkort, Carolin and Karzel, Marek and Zhukov, Evgeny A. and Evers, Eiko and Belykh, Vasilii V. and Bayer, Manfred},
  year = {2024},
  month = jul,
  journal = {ACS Photonics},
  volume = {11},
  number = {7},
  pages = {2757--2769},
  publisher = {American Chemical Society},
  doi = {10.1021/acsphotonics.4c00637},
  urldate = {2025-11-10}
}

@article{lampelNuclearDynamicPolarization1968,
  title = {Nuclear {{Dynamic Polarization}} by {{Optical Electronic Saturation}} and {{Optical Pumping}} in {{Semiconductors}}},
  author = {Lampel, Georges},
  year = {1968},
  month = mar,
  journal = {Phys. Rev. Lett.},
  volume = {20},
  number = {10},
  pages = {491--493},
  publisher = {American Physical Society},
  doi = {10.1103/PhysRevLett.20.491},
  urldate = {2025-11-05}
}

@article{Leppennen2022,
  title = {Optical Measurement of Electron Spins in Quantum Dots: Quantum {{Zeno}} Effects},
  author = {Leppenen, N. V. and Smirnov, D. S.},
  year = {2022},
  journal = {Nanoscale},
  volume = {14},
  number = {36},
  pages = {13284--13291},
  publisher = {The Royal Society of Chemistry},
  doi = {10.1039/D2NR01241C}
}

@article{lyubinskiy07,
  title = {Spin Dynamics in the Regime of Hopping Conductivity},
  author = {Lyubinskiy, I. S. and Dmitriev, A. P. and Kachorovskii, V. {\relax Yu}.},
  year = {2007},
  journal = {JETP Lett.},
  volume = {85},
  pages = {55}
}

@article{Merkulov2002,
  title = {Electron Spin Relaxation by Nuclei in Semiconductor Quantum Dots},
  author = {Merkulov, I. A. and Efros, {\relax Al}. L. and Rosen, M.},
  year = {2002},
  month = apr,
  journal = {Phys. Rev. B},
  volume = {65},
  number = {20},
  pages = {205309},
  publisher = {American Physical Society},
  doi = {10.1103/PhysRevB.65.205309}
}

@article{mikhailovElectronHoleSpin2018,
  title = {Electron and Hole Spin Relaxation in {{InP-based}} Self-Assembled Quantum Dots Emitting at Telecom Wavelengths},
  author = {Mikhailov, A. V. and Belykh, V. V. and Yakovlev, D. R. and Grigoryev, P. S. and Reithmaier, J. P. and Benyoucef, M. and Bayer, M.},
  year = {2018},
  month = nov,
  journal = {Phys. Rev. B},
  volume = {98},
  number = {20},
  pages = {205306},
  publisher = {American Physical Society},
  doi = {10.1103/PhysRevB.98.205306},
  urldate = {2025-11-10}
}

@article{MX2_Avdeev,
  title = {Hyperfine Interaction in Atomically Thin Transition Metal Dichalcogenides},
  author = {Avdeev, I. D. and Smirnov, D. S.},
  year = {2019},
  journal = {Nanoscale Adv.},
  volume = {1},
  number = {7},
  pages = {2624},
  publisher = {RSC},
  doi = {10.1039/C8NA00360B}
}

@article{nekrasovEffectNuclearQuadrupole2019,
  title = {Effect of Nuclear Quadrupole Interaction on Spin Beats in Photoluminescence Polarization Dynamics of Charged Excitons in {{InP}}/({{In}},{{Ga}}){{P}} Quantum Dots},
  author = {Nekrasov, S. V. and Akimov, I. A. and Kusrayev, {\relax Yu}. G. and Yakovlev, D. R. and Bayer, M.},
  year = {2019},
  month = dec,
  journal = {Phys. Rev. B},
  volume = {100},
  number = {23},
  pages = {235415},
  publisher = {American Physical Society},
  doi = {10.1103/PhysRevB.100.235415},
  urldate = {2025-11-10}
}

@article{noise_OMAR,
  title = {Electric Current Noise in Mesoscopic Organic Semiconductors Induced by Nuclear Spin Fluctuations},
  author = {Smirnov, D. S. and Shumilin, A. V.},
  year = {2021},
  month = may,
  journal = {Phys. Rev. B},
  volume = {103},
  number = {19},
  pages = {195440},
  publisher = {American Physical Society},
  doi = {10.1103/PhysRevB.103.195440}
}

@article{noise-trions,
  title = {Spin Noise Spectroscopy of a Single Quantum Well Microcavity},
  author = {Poltavtsev, S. V. and Ryzhov, I. I. and Glazov, M. M. and Koz\-lov, G. G. and Zapasskii, V. S. and Kavokin, A. V. and Lagoudakis, P. G. and Smirnov, D. S. and Ivchenko, E. L.},
  year = {2014},
  month = feb,
  journal = {Phys. Rev. B},
  volume = {89},
  number = {8},
  pages = {081304},
  publisher = {American Physical Society},
  doi = {10.1103/PhysRevB.89.081304}
}

@book{OptOr,
  title = {Optical Orientation},
  editor = {Meier, F. and Zakharchenya, B. P.},
  year = {1984},
  publisher = {North Holland, Amsterdam}
}

@article{petrov08,
  title = {Effect of Thermal Annealing on the Hyperfine Interaction in {{InAs}}/{{GaAs}} Quantum Dots},
  author = {Petrov, M. {\relax Yu}. and Ignatiev, I. V. and Poltavtsev, S. V. and Greilich, A. and Bauschulte, A. and Yakovlev, D. R. and Bayer, M.},
  year = {2008},
  month = jul,
  journal = {Phys. Rev. B},
  volume = {78},
  number = {4},
  pages = {045315},
  publisher = {American Physical Society},
  doi = {10.1103/PhysRevB.78.045315}
}

@article{PhysRevA.36.5543,
  title = {Quantum-Mechanical Model for Continuous Position Measurements},
  author = {Caves, C. M. and Milburn, G. J.},
  year = {1987},
  month = dec,
  journal = {Phys. Rev. A},
  volume = {36},
  number = {12},
  pages = {5543},
  publisher = {American Physical Society},
  doi = {10.1103/PhysRevA.36.5543}
}

@article{PhysRevA.84.030101,
  title = {Measurement of Noncommuting Spin Components Using Spin-Orbit Interaction},
  author = {Sokolovski, D. and Sherman, E. {\relax Ya}.},
  year = {2011},
  month = sep,
  journal = {Phys. Rev. A},
  volume = {84},
  number = {3},
  pages = {030101(R)},
  publisher = {American Physical Society},
  doi = {10.1103/PhysRevA.84.030101}
}

@article{PhysRevB.100.075409,
  title = {Universal Power Law Decay of Spin Polarization in Double Quantum Dot},
  author = {Mantsevich, V. N. and Smirnov, D. S.},
  year = {2019},
  month = aug,
  journal = {Phys. Rev. B},
  volume = {100},
  number = {7},
  pages = {075409},
  publisher = {American Physical Society},
  doi = {10.1103/PhysRevB.100.075409}
}

@article{PhysRevB.103.045413,
  title = {Quantum {{Zeno}} Effect and Quantum Nondemolition Spin Measurement in a Quantum Dot--Micropillar Cavity in the Strong Coupling Regime},
  author = {Leppenen, N. V. and Lanco, L. and Smirnov, D. S.},
  year = {2021},
  month = jan,
  journal = {Phys. Rev. B},
  volume = {103},
  number = {4},
  pages = {045413},
  publisher = {American Physical Society},
  doi = {10.1103/PhysRevB.103.045413},
  shorthand = {A15}
}

@article{PhysRevB.64.125316,
  title = {Spin-Flip Transitions between {{Zeeman}} Sublevels in Semiconductor Quantum Dots},
  author = {Khaetskii, A. V. and Nazarov, Y. V.},
  year = {2001},
  month = sep,
  journal = {Phys. Rev. B},
  volume = {64},
  number = {12},
  pages = {125316},
  publisher = {American Physical Society},
  doi = {10.1103/PhysRevB.64.125316}
}

@article{PhysRevB.66.161318,
  title = {Spin Relaxation in Quantum Dots},
  author = {Woods, L. M. and Reinecke, T. L. and {Lyanda-Geller}, Y.},
  year = {2002},
  month = oct,
  journal = {Phys. Rev. B},
  volume = {66},
  number = {16},
  pages = {161318(R)},
  publisher = {American Physical Society},
  doi = {10.1103/PhysRevB.66.161318}
}

@article{PhysRevB.98.125306,
  title = {Theory of Spin Inertia in Singly Charged Quantum Dots},
  author = {Smirnov, D. S. and Zhukov, E. A. and Kirstein, E. and Yakovlev, D. R. and Reuter, D. and Wieck, A. D. and Bayer, M. and Greilich, A. and Glazov, M. M.},
  year = {2018},
  month = sep,
  journal = {Phys. Rev. B},
  volume = {98},
  number = {12},
  pages = {125306},
  publisher = {American Physical Society},
  doi = {10.1103/PhysRevB.98.125306},
  shorthand = {A3}
}

@article{PhysRevLett.134.016201,
  title = {Cooling and Heating Nuclear Spins by Strongly Localized Electrons},
  author = {Smirnov, D. S. and Kavokin, K. V.},
  year = {2025},
  month = jan,
  journal = {Phys. Rev. Lett.},
  volume = {134},
  pages = {016201},
  publisher = {American Physical Society},
  doi = {10.1103/PhysRevLett.134.016201},
  shorthand = {A4}
}

@article{PhysRevLett.88.256801,
  title = {Manipulation of the Spin Memory of Electrons in {{n}}-{{GaAs}}},
  author = {Dzhioev, R. I. and Korenev, V. L. and Merkulov, I. A. and Zakharchenya, B. P. and Gammon, D. and Efros, {\relax Al}. L. and Katzer, D. S.},
  year = {2002},
  month = jun,
  journal = {Phys. Rev. Lett.},
  volume = {88},
  number = {25},
  pages = {256801},
  publisher = {American Physical Society},
  doi = {10.1103/PhysRevLett.88.256801}
}

@article{PhysRevLett.94.116601,
  title = {Direct Observation of the Electron Spin Relaxation Induced by Nuclei in Quantum Dots},
  author = {Braun, P.-F. and Marie, X. and Lombez, L. and Urbaszek, B. and Amand, T. and Renucci, P. and Kalevich, V. K. and Kavokin, K. V. and Krebs, O. and Voisin, P. and Masumoto, Y.},
  year = {2005},
  month = mar,
  journal = {Phys. Rev. Lett.},
  volume = {94},
  number = {11},
  pages = {116601},
  publisher = {American Physical Society},
  doi = {10.1103/PhysRevLett.94.116601}
}

@article{PRC,
  title = {Spin Polarization Recovery and {{Hanle}} Effect for Charge Carriers Interacting with Nuclear Spins in Semiconductors},
  author = {Smirnov, D. S. and Zhukov, E. A. and Yakovlev, D. R. and Kirstein, E. and Bayer, M. and Greilich, A.},
  year = {2020},
  month = dec,
  journal = {Phys. Rev. B},
  volume = {102},
  number = {23},
  pages = {235413},
  publisher = {American Physical Society},
  doi = {10.1103/PhysRevB.102.235413},
  shorthand = {A1}
}

@article{PRESILLA199695,
  title = {Measurement Quantum Mechanics and Experiments on Quantum Zeno Effect},
  author = {Presilla, C. and Onofrio, R. and Tambini, U.},
  year = {1996},
  journal = {Ann. Phys.},
  volume = {248},
  number = {1},
  pages = {95},
  issn = {0003-4916},
  doi = {10.1006/aphy.1996.0052}
}

@article{rautertOpticalOrientationAlignment2019,
  title = {Optical Orientation and Alignment of Excitons in Direct and Indirect Band Gap ({{In}},{{Al}}){{As}}/{{AlAs}} Quantum Dots with Type-{{I}} Band Alignment},
  author = {Rautert, J. and Shamirzaev, T. S. and Nekrasov, S. V. and Yakovlev, D. R. and Klenovsk{\'y}, P. and Kusrayev, {\relax Yu}. G. and Bayer, M.},
  year = {2019},
  month = may,
  journal = {Phys. Rev. B},
  volume = {99},
  number = {19},
  pages = {195411},
  publisher = {American Physical Society},
  doi = {10.1103/PhysRevB.99.195411},
  urldate = {2025-11-10}
}

@article{Shumilin2015,
  title = {Kinetic Equations for Hopping Transport and Spin Relaxation in a Random Magnetic Field},
  author = {Shumilin, A. V. and Kabanov, V. V.},
  year = {2015},
  month = jul,
  journal = {Phys. Rev. B},
  volume = {92},
  number = {1},
  pages = {014206},
  publisher = {American Physical Society},
  doi = {10.1103/PhysRevB.92.014206}
}

@article{shumilinMicroscopicTheoryOrganic2020,
  title = {Microscopic Theory of Organic Magnetoresistance Based on Kinetic Equations for Quantum Spin Correlations},
  author = {Shumilin, A. V.},
  year = {2020},
  month = apr,
  journal = {Phys. Rev. B},
  volume = {101},
  number = {13},
  pages = {134201},
  publisher = {American Physical Society},
  doi = {10.1103/PhysRevB.101.134201},
  urldate = {2025-10-02}
}

@article{smirnovSpinPolarizationRecovery2020,
  title = {Spin Polarization Recovery and {{Hanle}} Effect for Charge Carriers Interacting with Nuclear Spins in Semiconductors},
  author = {Smirnov, D. S. and Zhukov, E. A. and Yakovlev, D. R. and Kirstein, E. and Bayer, M. and Greilich, A.},
  year = {2020},
  month = dec,
  journal = {Phys. Rev. B},
  volume = {102},
  number = {23},
  pages = {235413},
  publisher = {American Physical Society},
  doi = {10.1103/PhysRevB.102.235413},
  urldate = {2025-11-05}
}

@article{sokolovNuclearSpinCooling2017,
  title = {Nuclear Spin Cooling by Helicity-Alternated Optical Pumping at Weak Magnetic Fields in \$n\$-{{GaAs}}},
  author = {Sokolov, P. S. and Petrov, M. {\relax Yu}. and Kavokin, K. V. and Kurdyubov, A. S. and Kuznetsova, M. S. and Cherbunin, R. V. and Verbin, S. {\relax Yu}. and Poletaev, N. K. and Yakovlev, D. R. and Suter, D. and Bayer, M.},
  year = {2017},
  month = nov,
  journal = {Phys. Rev. B},
  volume = {96},
  number = {20},
  pages = {205205},
  publisher = {American Physical Society},
  doi = {10.1103/PhysRevB.96.205205},
  urldate = {2025-11-10}
}

@article{UspFizNauk2021,
  title = {Theory of Optically Detected Spin Noise in Nanosystems},
  author = {Smirnov, D. S. and Mantsevich, V. N. and Glazov, M. M.},
  year = {2021},
  journal = {Phys. Usp.},
  volume = {64},
  number = {9},
  pages = {923--946},
  doi = {10.3367/UFNe.2020.10.038861}
}

@article{yugova09,
  title = {Pump-Probe {{Faraday}} Rotation and Ellipticity in an Ensemble of Singly Charged Quantum Dots},
  author = {Yugova, I. A. and Glazov, M. M. and Ivchenko, E. L. and Efros, {\relax Al}. L.},
  year = {2009},
  journal = {Phys. Rev. B},
  volume = {80},
  number = {104436},
  pages = {104436},
  publisher = {APS},
  doi = {10.1103/PhysRevB.80.104436}
}

@article{Zeno_exp,
  title = {Tuning the Nuclei-Induced Spin Relaxation of Localized Electrons by the Quantum {{Zeno}} and Anti-{{Zeno}} Effects},
  author = {Nedelea, V. and Leppenen, N. V. and Evers, E. and Smirnov, D. S. and Bayer, M. and Greilich, A.},
  year = {2023},
  month = sep,
  journal = {Phys. Rev. Res.},
  volume = {5},
  number = {3},
  pages = {L032032},
  publisher = {American Physical Society},
  doi = {10.1103/PhysRevResearch.5.L032032},
  shorthand = {A17}
}

@article{Zhukov07,
  title = {Spin Coherence of a Two-Dimensional Electron Gas Induced by Resonant Excitation of Trions and Excitons in {{CdTe}}/({{Cd}},{{Mg}}){{Te}} Quantum Wells},
  author = {Zhukov, E. A. and Yakovlev, D. R. and Bayer, M. and Glazov, M. M. and Ivchenko, E. L. and Karczewski, G. and Wojtowicz, T. and Kossut, J.},
  year = {2007},
  month = nov,
  journal = {Phys. Rev. B},
  volume = {76},
  number = {20},
  pages = {205310},
  publisher = {American Physical Society},
  doi = {10.1103/PhysRevB.76.205310}
}

@InBook{yafet63,
  Title                    = {$g$-factors and spin-lattice relaxation of conduction electrons},
  Author                   = {Y. Yafet},
  Editor                   = {F. Seitz and D. Turnbull},
  Volume                   = {14},
  Pages                    = {1},
  Publisher                = {Academic, New-York},
  Year                     = {1963},

  Booktitle                = {Solid State Physics}
}

@Article{LossPRA98,
  Title                    = {Quantum computation with quantum dots},
  Author                   = {Loss, D. and DiVincenzo, D. P.},
  Journal                  = {Phys. Rev. A},
  Year                     = {1998},

  Month                    = {Jan},
  Pages                    = {120--126},
  Volume                   = {57},
  Doi                      = {10.1103/PhysRevA.57.120},
  Issue                    = {1},
  Publisher                = {American Physical Society},
  Url                      = {http://link.aps.org/doi/10.1103/PhysRevA.57.120}
}

@Article{press08,
  Title                    = {{Complete quantum control of a single quantum dot spin using ultrafast optical pulses}},
  Author                   = {Press, D. and Ladd, T. D. and Zhang, B. and Yamamoto, Y.},
  Journal                  = {Nature (London)},
  Year                     = {2008},
  Pages                    = {218},
  Volume                   = {456}
}

@Article{PhysRevB.101.075412,
  Title                    = {{Electron-nuclei interaction in the $X$ valley of (In,Al)As/AlAs quantum dots}},
  Author                   = {Kuznetsova, M. S. and Rautert, J. and Kavokin, K. V. and Smirnov, D. S. and Yakovlev, D. R. and Bakarov, A. K. and Gutakovskii, A. K. and Shamirzaev, T. S. and Bayer, M.},
  Journal                  = {Phys. Rev. B},
  Year                     = {2020},

  Month                    = {Feb},
  Pages                    = {075412},
  Volume                   = {101},

  Doi                      = {10.1103/PhysRevB.101.075412},
  Issue                    = {7},
  Numpages                 = {9},
  Publisher                = {American Physical Society},
  Url                      = {https://link.aps.org/doi/10.1103/PhysRevB.101.075412}
}

@Article{schulten,
  Title                    = {Semiclassical description of electron spin motion in radicals including the effect of electron hopping},
  Author                   = {K. Schulten and P. G. Wolynes},
  Journal                  = {J. Chem. Phys.},
  Year                     = {1978},
  Number                   = {7},
  Pages                    = {3292},
  Volume                   = {68},

  Doi                      = {10.1063/1.436135},
  Publisher                = {AIP},
  Url                      = {http://link.aip.org/link/?JCP/68/3292/1}
}

@Article{PhysRevResearch.1.033189,
  Title                    = {{Spin inertia and polarization recovery in quantum dots: Role of pumping strength and resonant spin amplification}},
  Author                   = {Schering, P. and Uhrig, G. S. and Smirnov, D. S.},
  Journal                  = {Phys. Rev. Research},
  Year                     = {2019},

  Month                    = {Dec},
  Pages                    = {033189},
  shorthand                = {A2},
  Volume                   = {1},

  Doi                      = {10.1103/PhysRevResearch.1.033189},
  Issue                    = {3},
  Numpages                 = {12},
  Publisher                = {American Physical Society},
  Url                      = {https://link.aps.org/doi/10.1103/PhysRevResearch.1.033189}
}

@Article{efros03,
  Title                    = {Optical readout and initialization of an electron spin in a single quantum dot},
  Author                   = {Shabaev, A. and Efros, Al. L. and Gammon, D. and Merkulov, I. A.},
  Journal                  = {Phys. Rev. B},
  Year                     = {2003},

  Month                    = {Nov},
  Number                   = {20},
  Pages                    = {201305},
  Volume                   = {68},

  Numpages                 = {4}
}

@Article{yugova12,
  Title                    = {Coherent spin dynamics of electrons and holes in semiconductor quantum wells and quantum dots under periodical optical excitation: Resonant spin amplification versus spin mode locking},
  Author                   = {I. A. Yugova and M. M. Glazov and D. R. Yakovlev and A. A. Sokolova and M. Bayer},
  Journal                  = {Phys. Rev. B},
  Year                     = {2012},
  Pages                    = {125304},
  Volume                   = {85}
}

@Article{LANE1983359,
  Title                    = {{Decay at early times: Larger or smaller than the golden rule?}},
  Author                   = {A. M. Lane},
  Journal                  = {Phys. Lett. A},
  Year                     = {1983},
  Number                   = {8},
  Pages                    = {359},
  Volume                   = {99},
  Doi                      = {https://doi.org/10.1016/0375-9601(83)90292-X},
  ISSN                     = {0375-9601},
  Url                      = {http://www.sciencedirect.com/science/article/pii/037596018390292X}
}
\end{document}